\documentclass[floatfix,%
 aip,
 amsmath,amssymb,
 reprint,%
]{revtex4-1}

\usepackage{graphicx}% Include figure files
\usepackage{dcolumn}% Align table columns on decimal point
\usepackage{bm}% bold math
\usepackage[utf8]{inputenc}
\usepackage[T1]{fontenc}
\usepackage{mathptmx}
\usepackage{etoolbox}

\usepackage{xr}
\makeatletter
\def\@email#1#2{%
 \endgroup
 \patchcmd{\titleblock@produce}
  {\frontmatter@RRAPformat}
  {\frontmatter@RRAPformat{\produce@RRAP{*#1\href{mailto:#2}{#2}}}\frontmatter@RRAPformat}
  {}{}
}%
\makeatother
\begin{document}

\preprint{}

\title[]{Surface Morphology Controls Charge Storage at the Electrified Pt–Water Interface}
% Force line breaks with \\
\author{Matthew T. Darby}
\affiliation{ 
Department of Chemistry, Faculty of Natural Sciences,Imperial College London, London%\\This line break forced with \textbackslash\textbackslash
}
\altaffiliation[]{Thomas Young Centre}%Lines break automatically or can be forced with \\
\author{Muhammad Saleh}%
% \email{Second.Author@institution.edu.}
\affiliation{ 
Theoretical physics of electrified liquid-solid interfaces,RUHR-Universit\"{a}t Bochum, 44801 Bochum, Germany%\\This line break forced with \textbackslash\textbackslash
}%

\author{Marialore Sulpizi}%
% \email{Second.Author@institution.edu.}
\affiliation{ 
Theoretical physics of electrified liquid-solid interfaces,RUHR-Universit\"{a}t Bochum, 44801 Bochum, Germany%\\This line break forced with \textbackslash\textbackslash
}%

\author{Clotilde S. Cucinotta$^{*,}$}
\email{c.cucinotta@imperial.ac.uk}
\affiliation{ 
Department of Chemistry, Faculty of Natural Sciences,Imperial College London, London%\\This line break forced with \textbackslash\textbackslash
}
\altaffiliation[]{Thomas Young Centre}%Lines break automatically or can be forced with \\

\date{\today}% It is always \today, today,

% %%%%%%%%%%%%%%%%%%%%%%%%%%%%%%%%%%%%%%%%%%%%%%%%%%%%%%%%%%%%%%%%%%%%%
% %% KEYWORD
% %%%%%%%%%%%%%%%%%%%%%%%%%%%%%%%%%%%%%%%%%%%%%%%%%%%%%%%%%%%%%%%%%%%%%
% \abbreviations{Platinum, electrochemistry, electrical double layer, fuel cells, density functional theory, HER, \emph{ab initio} molecular dynamics}
% \keywords{American Chemical Society, \LaTeX}

\begin{abstract}
Platinum step edges dominate electrocatalytic activity in fuel cells and electrolysers, yet their atomistic electrochemical behaviour remains poorly understood. Here, we employ ab initio molecular dynamics under controlled electrode potentials to model a realistic stepped Pt–water interface incorporating experimentally observed (111)$\times$(111) and (111)$\times$(100) edge motifs. This allows us to resolve, for the first time, the site-specific structure, charge distribution, and electrostatics of the electric double layer at a nanostructured Pt surface.

We find that differential capacitance near the potential of zero charge (PZC) arises almost entirely from potential-dependent chemisorption of water on flat (111) terraces. In contrast, step edges are saturated with chemisorbed water even below the PZC and thus do not contribute to the capacitance. Instead, edges accumulate excess positive charge and exhibit a locally elevated electrostatic potential, as revealed by spatially resolved macroscopic potential profiles. This electrostatic asymmetry implies a greater barrier for electron accumulation at step sites compared to terraces, consistent with enhanced charge localisation and reactivity. 

Finally, the higher in energy d-band centre and sharper projected density of states at edge atoms further support their role as active, positively charged centres. Together, these results provide a mechanistic explanation for the observed experimental shift of the PZC with step density and establish a predictive framework for understanding and optimising interfacial charging in nanostructured Pt electrocatalysts.

\end{abstract}

\maketitle

% \begin{quotation}
% The ``lead paragraph'' is encapsulated with the \LaTeX\ 
% \verb+quotation+ environment and is formatted as a single paragraph before the first section heading. 
% (The \verb+quotation+ environment reverts to its usual meaning after the first sectioning command.) 
% Note that numbered references are allowed in the lead paragraph.
% %
% The lead paragraph will only be found in an article being prepared for the journal \textit{Chaos}.
% \end{quotation}

% INTRODUCTION
%%%%%%%%%%%%%%%%%%%%%%%%%%%%%%%%%%%%
\section{\label{sec:level1}Introduction}

Platinum (Pt) and its alloys are widely recognised as benchmark electrocatalytic systems as a result of their high activity, selectivity, and stability under harsh electrochemical conditions. Pt-based catalysts are central to fuel cells and acidic water electrolysers, in which they drive the hydrogen evolution and oxidation reactions (HER and HOR). However, their high cost and limited availability remain major barriers to broader deployment\cite{xun2025sustainable}. Although alternative materials are under investigation, few match the catalytic performance of Pt, particularly in acidic environments.
A key challenge in catalyst optimisation is the lack of atomistic insight into their nanoscale electrochemical interfaces. Traditional models of the electrical double layer (EDL), such as those developed by Helmholtz, Gouy, Chapman, and Stern, offer macroscopic insight but cannot resolve the structure and charge redistribution at the atomic scale\cite{schott2024assess}.

These continuum models treat the electric double layer (EDL) formation as a purely electrostatic phenomenon, an assumption now being challenged by recent theoretical advances\cite{Darby2022,todorova2024first}.
A major milestone has been the application of \textit{ab initio} molecular dynamics (AIMD) to investigate Pt–water interfaces, providing a much clearer understanding of the EDL structure at the flat Pt(111) surface and enabling more direct comparison with experimental observations. AIMD explicitly captures phenomena such as interfacial polarisation, and enables atomistic resolution of the EDL structure and dynamics. For the idealised Pt(111) surface, AIMD simulations reveal that water molecules directly contacting the surface become partially charged and chemisorb \cite{Raffone2025,Khatib2021,Surendralal2021, Le2020}, thereby acting more like ionic species than neutral dipoles. This chemisorbed water contributes substantially to interfacial screening and differential capacitance\cite{Raffone2025,Surendralal2021}. Moreover, around the potental of zero charge the coverage of chemisorbed water correlates linearly with metal capacitance and potential increase\cite{Khatib2021}, although this relationship breaks down at higher potentials, due to water coverage saturation
\cite{Le2021},
and at lower potentials, due to H adsorption\cite{Surendralal2021,Raffone2025}. 
The orientation of water molecules in the 1\textsuperscript{st} and 2\textsuperscript{nd} interfacial layers can be broadly classified into three distinct motifs:  
(i) "Chemisorbed" water, charged and lying nearly flat on the Pt surface with oxygen coordinated to the metal;  
(ii) "H-down" water, in which one O–H bond points toward the surface, forming a weak covalent interaction with Pt;  
(iii) "Bridging" water, which hydrogen-bonds to both the 1\textsuperscript{st} and 2\textsuperscript{nd} layers, as well as to the bulk water phase, mediating vertical and lateral connectivity across the interface.\cite{Khatib2021,Le2020,Sakong2020,Surendralal2021}

However, real electrocatalysts are not atomically flat. Practical electrodes are nanostructured, with terraces, step edges, and undercoordinated sites that modulate interfacial behaviour\cite{Li2021a,Gomez2000,Kolb2016}. Under operating conditions, even initially pristine Pt(111) electrodes undergo electrochemical roughening, as revealed by in situ EC-STM. Repeated cyclic voltammetry induces the formation of nanoscale topographies comprising extended terraces, three-atom-deep step edges, and shallow, narrow pits that evolve progressively over tens to hundreds of cycles.\cite{Jacobse2019} This morphological evolution directly impacts interfacial chemistry. 
Indeed, in situ studies report a correlation between increasing densities of (111)$\times$(100) and (111)$\times$(111) step sites and enhanced current densities for the hydrogen evolution and oxidation reactions (HER/HOR).\cite{Jacobse2019,Zalitis2017,Gomez2000} These findings support the long-standing hypothesis that catalytic activity on Pt is dominated by undercoordinated edges, rather than flat terraces, 
highligting  the need for molecular-level models that resolve such nanoscale features.
In early efforts to rationalise the behaviour of water on non-flat Pt electrodes, Koper and coworkers performed Density Functional Theory (DFT) calculations to explore water adsorption near step edges on high index Pt(533) surface, with a simplified model for the electrolyte \cite{Kolb2016}. It was found that water forms stable hexagonal or pentagonal rings at step edges, with predominant H-down orientations. At higher water coverages, multiple near-degenerate structures form along the step, highlighting its role in anchoring water. Beyond static DFT approaches  Chen\textit{ et al.} have utilised AIMD to explore water structuring on several stepped Pt/water interfaces at the potential of zero charge (PZC) \cite{Chen2022}. It was found that water preferentially chemisorbs at step sites, with terrace adsorption occurring only at very low step densities. Stepped surfaces also exhibit lower PZCs than flat Pt(111), due to the difference in work function for these orientation. A recent study from Wang \textit{et al.} \cite{Wang2024} used machine learning molecular dynamics to investigate water behavior at the stepped Pt(211)/water interface. The results reveal both chemisorbed and physisorbed water types, along with three unique water pairs not seen on flat Pt surfaces that have been identified as potentially being important for water dissociation.\\
However, existing models often consider idealised stepped surfaces with high step densities and narrow terraces, neglecting the nanoscale terrace–edge architecture observed under electrochemical roughening conditions. Moreover, most simulations are conducted at fixed charge or at the PZC, without explicit control of electrode potential. \\
In this work, we build on these recent dynamic EDL studies by developing a more realistic Pt–electrolyte interface model that explicitly incorporates nanoscale terraces and undercoordinated step edges. We investigate how these interfacial nanostructures shape EDL behaviour, revealing the key factors that govern water adsorption, charge redistribution, potential drop, capacitance, and reactivity.

Our findings provide direct atomistic insight into how surface morphology governs in interfacial charge storage. By linking nanoscale geometry to the structure and capacitance of the electric double layer (EDL), we establish a mechanistic basis for observed changes in Pt–water interfacial nanostructuring. Such advances in nanoscopic understanding of the EDL can inform the design of more efficient and resource-optimised fuel cells and electrolysers.

% COMPUTATIONAL DETAIL SECTION
%%%%%%%%%%%%%%%%%%%%%%%%%%%%%%%%%%%%
\section{\label{sec:computational detail}Computational Details}
\subsection{\label{sec:model system}Model system and electrode potential referencing}
\subsubsection{\label{sec:system description} System description}

%FIGURE1
%%%%%%%%%%%%%%%%%%%%%%%%%%%%%%%%%%%%
\begin{figure*}
\includegraphics[width=0.70\textwidth]{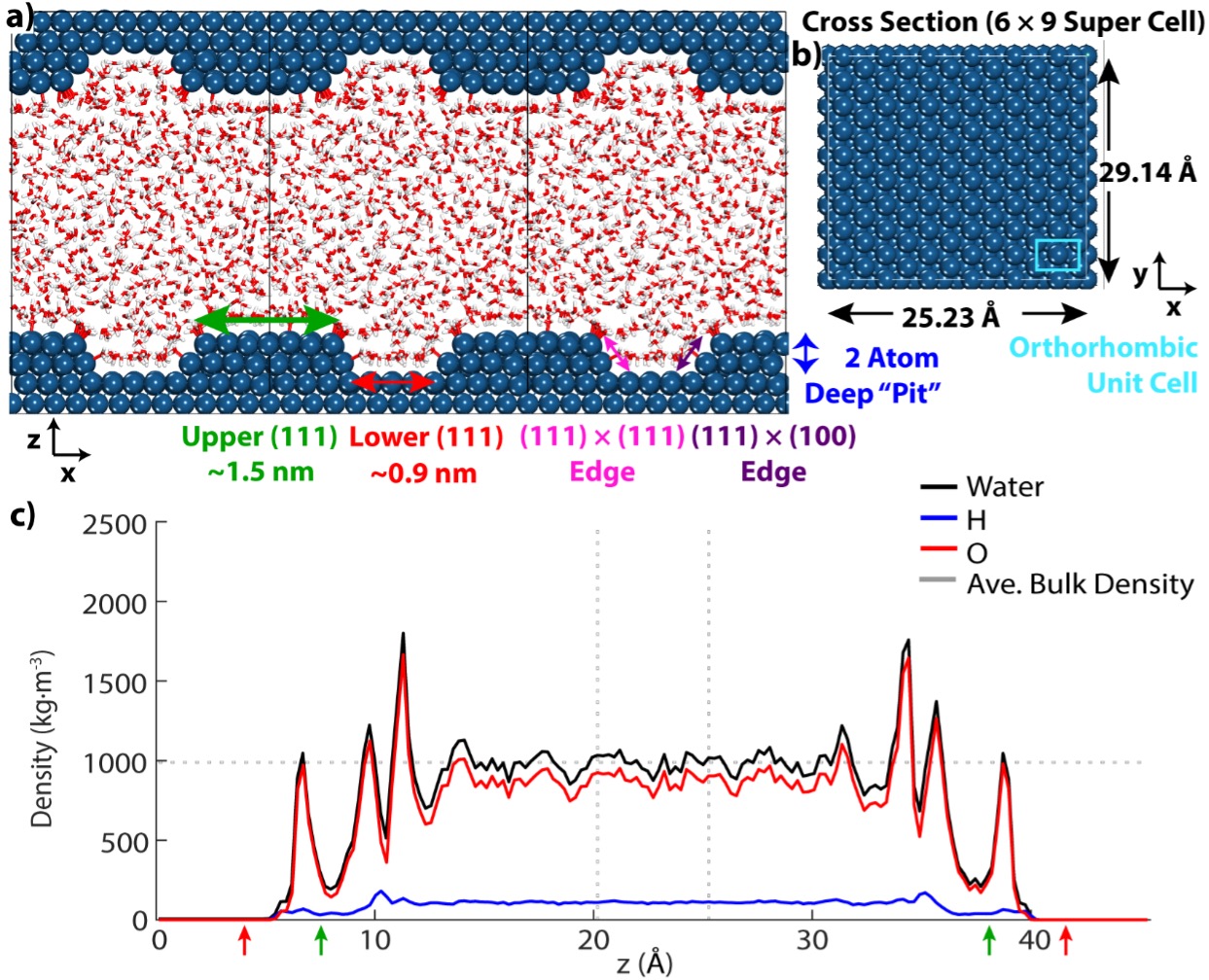}% Here is how to import EPS art
\caption{\label{fig:figure1}Our model for the Interface between Pt and a pure aqueous electrolyte. (a) Representation of three replicas along the $x$-axis of the used supercell. Pt, O and H atoms are represented in blue, red and white respectively. Periodic boundaries are marked with black lines, and the cell is oriented with the $y$-axis directed into the page. The upper (111) terrace, lower (111) terrace, angled (111) facet, and angled (100) facet are indicated by red, green, magenta, and purple arrows, respectively.  (b) Cross section of the supercell in the $xy$ plane showing the $6 \times 9$  replica of the orthorhombic unitary supercell (marked in cyan).  (c) Planar-averaged mass density distribution along $z$ for the system solvated in a pure aqueous electrolyte. Distributions of water molecules, H atoms, and O atoms are shown in black, blue, and red, respectively. The average bulk density is indicated by the horizontal grey line and is computed within the central 5 Å of the cell, as indicated by the vertical grey lines.}
\end{figure*}

We developed a unique, large-scale atomistic model of the Pt--electrolyte interface that reproduces key features of the experimental cyclic voltammetry and electrochemical scanning tunnelling microscopy (EC-STM) observations of Jacobse \textit{et al.}\,\cite{Jacobse2019} (Fig. \ref{fig:figure1}). This model enables direct comparison within the same simulation cell of the behaviour of terrace and edge sites under electrochemical bias.

Our model consists of a stepped Pt slab containing two distinct edge motifs, (111)$\times$(111) and (111)$\times$(100), formed at the junctions between (111) terraces and (111) or (100) facets. These edges are separated by two atomically flat terraces approximately 0.9~nm and 1.5~nm wide (Fig.~\ref{fig:figure1}a). The slab was constructed by replicating a $6\times9$ orthorhombic Pt(111) unit cell to form a structure with dimensions $\mathbf{a} = 29.14$~\AA{}, $\mathbf{b} = 25.23$~\AA{}, and $\mathbf{c} \sim 45.4$~\AA{}, with $\mathbf{c}$ normal to the surface. 
The upper terrace spans seven Pt atom rows ($\sim$1.5~nm) and descends into a stepped pit defined by inclined (111) and (100) facets at approximately 65$^\circ$ and 70$^\circ$, respectively. These sites are structurally and chemically distinct, and representative of the undercoordinated edges observed during electrochemical roughening of Pt(111) surfaces.\cite{Jacobse2019}.

The resulting slab contains 702 Pt atoms in an FCC lattice (lattice constant 3.96~\AA{}), arranged into eight atomic layers except within the central pit, where it narrows to four layers.  The central two Pt layers were constrained during simulation to reproduce bulk Pt conditions. The system was solvated with 714 water molecules and an adjustable number of H$^+$ and F$^-$ ions. The vertical dimension of the cell was optimised to yield a water density of approximately 1~g\,cm$^{-3}$ at 300~K in the centre of the electrolyte region (Fig.~\ref{fig:figure1}c).

\subsubsection{\label{sec:electrode pot descriptor} Electrode potential control and referencing}

We modelled a Pt--1.8~mol\,L$^{-1}$ HF aqueous solution interface at four electrode potentials near the potential of zero charge (PZC), using the ion imbalance method, which controls the electrode potential by introducing an imbalance in the number of ions in solution and allowing the electrode to charge in response (see SI section~S1 for details). Each cell includes a central bulk-like aqueous region that defines the same chemical state and serves as a reference for aligning electrostatic potential profiles.

The potential drop, $\Delta \psi$, is evaluated with respect to the electrostatic potential in the bulk electrolyte (see SI section~S2 for details). Specifically, the average Hartree potential profile along the $z$-direction was calculated for each trajectory snapshot and subsequently averaged over the production trajectory to obtain $V_{\mathrm{wat}}$ and $V_{\mathrm{m}}$. The metal potential $V_{\mathrm{m}}$ was determined from the central two atomic layers of the Pt slab, with a macroscopic average using a $z$-averaging window %of $\pm d_{\mathrm{Pt\text{-}Pt}}/2$ from the plane of each layer, where 
equal to $d_{\mathrm{Pt\text{-}Pt}}$, the interlayer spacing in bulk Pt along [111]. The electrolyte potential $V_{\mathrm{wat}}$ was obtained by averaging the $x$--$y$ planar Hartree potential over a 6~\AA~region centred at the midpoint along $z$ between the slab and its periodic image. Because the number of water molecules in this region is far from the thermodynamic limit, the instantaneous potential profiles exhibit significant fluctuations; trajectory-averaging after profile calculation mitigates this noise.

The relative electrode potential was then defined as
\begin{equation}
\Delta V = \Delta \psi - \Delta \psi_{\mathrm{PZC}},
\end{equation}
with $\Delta \psi_{\mathrm{PZC}}$ the value for the Pt--water system. This defines a potential window from $-0.01$~V to $+0.19$~V relative to the PZC. \\ To express $V$ on the reversible hydrogen electrode (RHE) scale, we align the reference system to the experimental PZC of stepped Pt(111), $V_{\mathrm{PZC}}^{\mathrm{exp}} = 0.53$~V vs.\ RHE for a cell with step density comparable to ours,\cite{Gomez2000} yielding
\begin{equation}
V = \Delta V + 0.53~\mathrm{V}.
\end{equation}
This referencing protocol is consistent with the alignment method introduced in Ref.~\citenum{Raffone2025}. The resulting electrode potentials span the range 0.52--0.71~V vs.\ RHE (see Table~\ref{tab:table2}).

Although $V$ depends on pH, all systems correspond to $\mathrm{pH} = 0$, set by the hydronium content in the bulk region. No pH correction is required.

\subsection{\label{sec:simulation protocol}Simulation Protocol}

We performed ab initio molecular dynamics (AIMD) simulations using the open-source code \textsc{CP2K}, version 8.1~\cite{Kuhne2020}. Atomic dynamics were propagated with the Car–Parrinello–like scheme developed by Kühne \textit{et al.}~\cite{Kuhne2007}. The simulations were carried out in the canonical (NVT) ensemble with a time step of 0.5 fs at 300 K.  After an initial 10 ps equilibration to ensure thermal and structural stability, we accumulated for each system 20–25 ps of production trajectories, providing statistical sampling of interfacial fluctuations. The charge localized on the electrode remained constant throughout this phase.

The electronic structure calculations were carried out within the generalised gradient approximation using the PBE exchange--correlation functional~\cite{Perdew1996}, and dispersion interactions were treated using Grimme’s third-generation DFT-D3 correction~\cite{Grimme2010}. A planewave cutoff energy of 300~Ry was used, and all calculations were performed at the $\Gamma$-point. Triple-zeta basis sets were used for all species, with added polarisation functions for F, O, and H. Core electrons were modelled using Goedecker--Teter--Hutter (GTH) pseudopotentials with valences of 1, 6, 7, and 18 for H, O, F, and Pt, respectively. 

%FIGURE2
%%%%%%%%%%%%%%%%%%%%%%%%%%%%%%%%%%%%
\begin{figure*}
\includegraphics[width=0.75\textwidth]{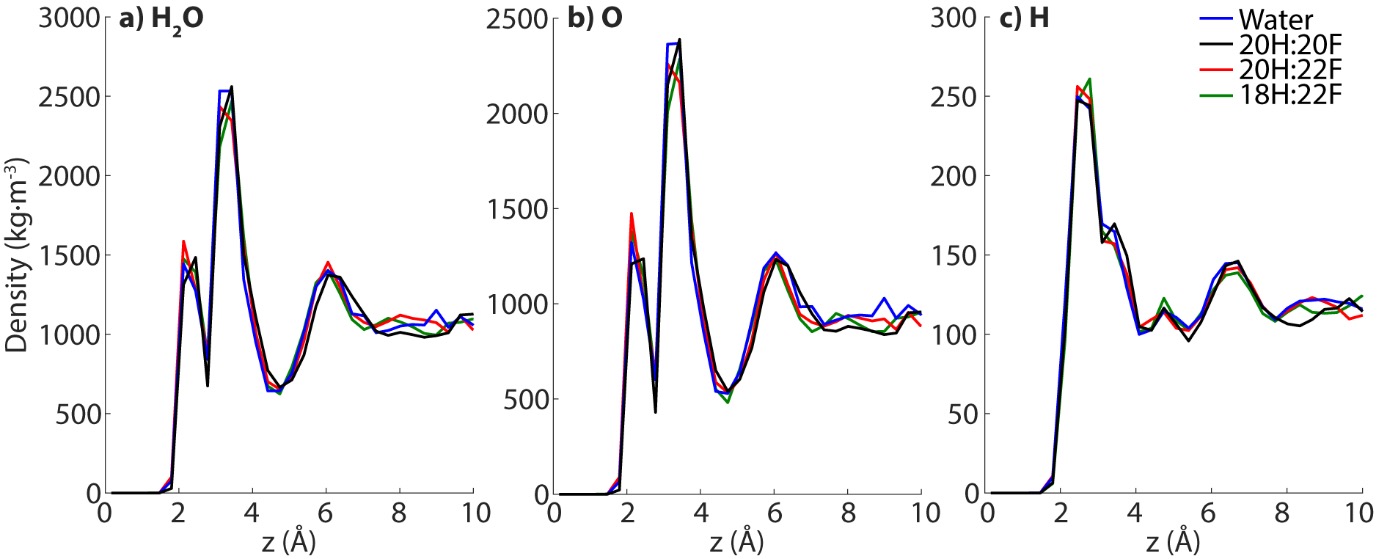}% Here is how to import EPS art
\caption{\label{fig:figure2}Topologically sensitive mass density distributions for a) water molecules, b) O atoms, and c) H atoms. Systems include pure water (blue) and three HF electrolytes with H:F ratios of 20:20 (black), 20:22 (red), and 18:22 (green). Distance $d$ is defined as the minimum distance to the surface plane}
\end{figure*}

The Always Stable Predictor Corrector (ASPC) method was used to propagate the wavefunction with an extrapolation order of zero and an initial line search step size of 0.01. A full single inverse preconditioner was employed before each SCF step, using an energy gap of $10^{-3}$~Hartree. The electronic structure was optimised via orbital transformation to a convergence threshold of $10^{-5}$~Hartree or a maximum of three self-consistent field (SCF) steps and corrected by a stochastic term following Kühne \textit{et al.}~\cite{Kuhne2007}, with a Langevin friction coefficient of $3.75 \times 10^{-4}$~fs$^{-1}$ to account for residual forces arising from incomplete SCF convergence.

% RESULT AND DISCUSSION
%%%%%%%%%%%%%%%%%%%%%%%%%%%%%%%%%%%%
\section{\label{sec:result}Results and Discussion}
\subsection{\label{structural-analysis}Double layer structure}

At electrified metal–electrolyte interfaces, charge transfer and polarisation drive pronounced structural ordering of interfacial water. On stepped Pt surfaces, this ordering exhibits clear site specificity. Water chemisorbs more readily at undercoordinated edge sites, reaching full saturation at potentials below the PZC, while terrace adsorption remains lower but increases linearly with potential\cite{Khatib2021}. Across all sites, chemisorbed molecules adopt a nearly flat geometry and form directional, zigzag hydrogen-bonded chains along edges. A more dinamically structured second water layer connects these chemisorbed molecules laterally and to the bulk, and near the edges, wraps around the saturated zigzag network to form a cylindrical shell. This shell encloses a low-density void above the chemisorbed layer and gives rise to an apparent third peak in the density profiles, though no structural third layer is structurally resolved. These features reveal a complex, topography-dependent interfacial structure inaccessible to planar models.

%FIGURE3
%%%%%%%%%%%%%%%%%%%%%%%%%%%%%%%%%%%%
\begin{figure}
\includegraphics[width=0.45\textwidth]{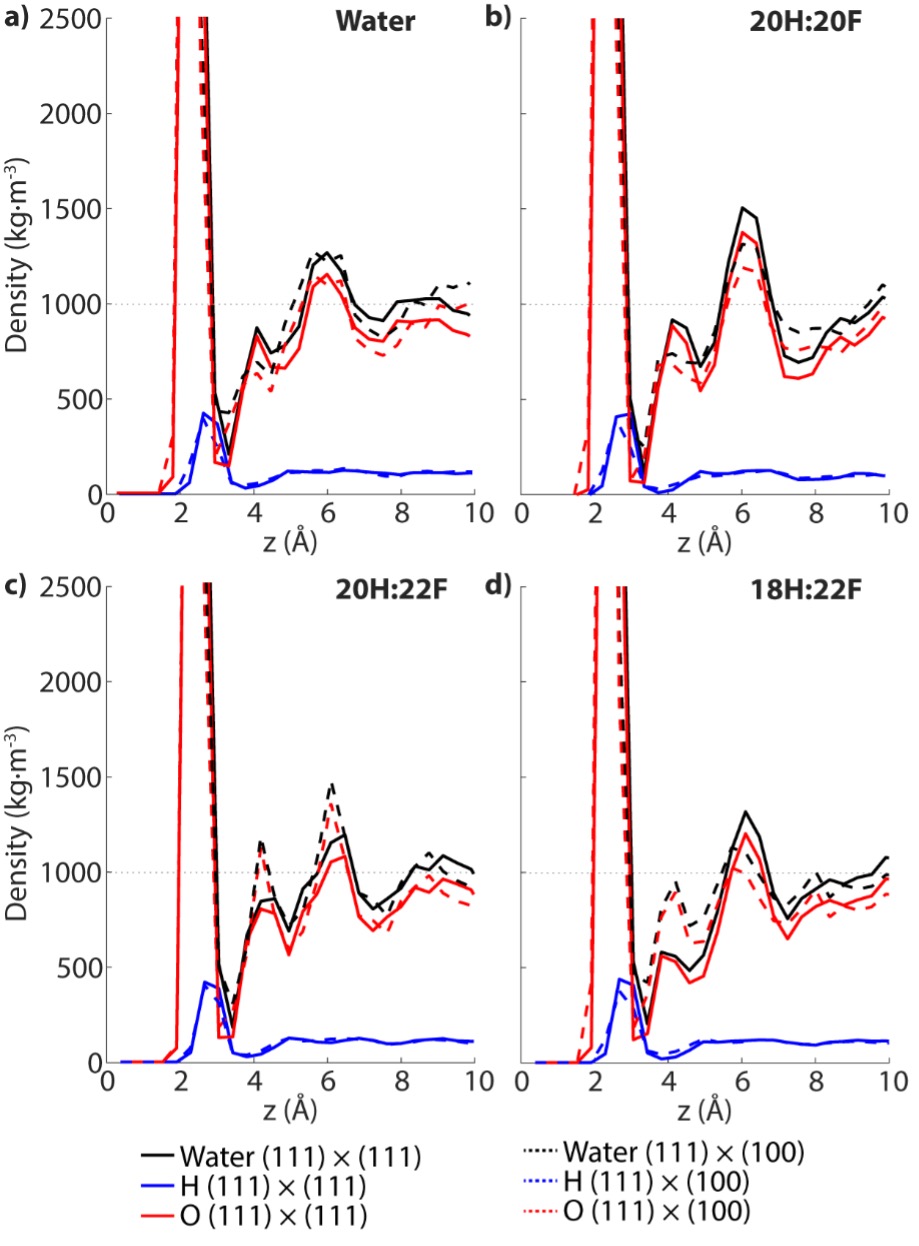}% Here is how to import EPS art
\caption{\label{fig:figure3} Cylindrical mass density distributions centred about the (111)$\times$(111) (solid lines) and (111)$\times$(100) (dashed lines) edges for water (black), oxygen (red), and hydrogen (blue). Panels show a) water-only, and b–d) HF solutions with H:F ratios of 20:20, 20:22, and 18:22, respectively. See supplementary material section 1 and 2 for details.}
\end{figure}

To quantify this structuring, we analysed the water oxygen mass density using three complementary metrics tailored to stepped morphologies (Figure S1) the planar mass density (PMD), the topologically sensitive distribution (TSD), and the cylindrical average distribution (CAD). The PMD averages density along the direction normal to the (111) terrace, capture the vertical layering (Figure~\ref{fig:figure1}c). The TSD resolves mass layering relative to the true atomic surface by projecting local distances onto reference planes fitted to the Pt topology (Figure~\ref{fig:figure2}). The CAD probes radial water structuring around edges by averaging density in cylindrical coordinates centred on edge rows (Figure~\ref{fig:figure3}). Together, these descriptors reveal the lateral and vertical anisotropies in interfacial water induced by nanoscale topography. Full computational details are given in the SI section~ S3. The quantitative analysis of these distributions, including the number of water molecules per interfacial layer, is summarised in Table~\ref{tab:table1}. A planar-averaged profile restricted to the upper (111) terrace was also computed to isolate terrace-specific features. The integration bounds are in this case defined by the edges of the cylindrical regions around the step sites.

%TABLE1
%%%%%%%%%%%%%%%%%%%%%%%%%%%%%%%%%%%%
\begin{table*}
\caption{\label{tab:table1}Analysis of integrated mass density distributions above specific regions of the Pt electrode surface—the full surface, the (111)$\times$(111) edge, the (111)$\times$(100) edge, and the upper (111) terrace—between the first, second, and third peaks identified in each profile. The oxygen coverage is reported in monolayers (ML), obtained by normalising the number of water molecules to the number of Pt atoms in each region. To account for topological differences, the topologically sensitive distribution (TSD) was used for the full surface, the cylindrical average distribution (CAD) for the step edges, and a restricted planar mass distribution (PMD) for the upper terrace. Coverages are shown for each electrode potential studied, along with the corresponding ion imbalance used to control the electrochemical bias.}

\begin{ruledtabular}
\begin{tabular}{lcccc}
System & Water (PZC) & 20H:20F & 20H:22F & 18H:22F \\
\hline
Potential (V vs RHE)   & 0.53 & 0.52 & 0.64 & 0.71 \\
Potential (V vs PZC)   & 0.00 & 0.01 & 0.11 & 0.18 \\
\hline
\multicolumn{5}{l}{1st Density Peak} \\
Full Surface            & 0.17 & 0.18 & 0.19 & 0.19 \\
(111)$\times$(111) Edge & 0.92 & 0.95 & 0.92 & 0.93 \\
(111)$\times$(100) Edge & 0.89 & 0.93 & 0.87 & 0.86 \\
(111) Terrace           & 0.06 & 0.06 & 0.07 & 0.09 \\
\hline
\multicolumn{5}{l}{2nd Density Peak} \\
Full Surface            & 0.56 & 0.58 & 0.58 & 0.57 \\
(111)$\times$(111) Edge & 0.36 & 0.47 & 0.50 & 0.27 \\
(111)$\times$(100) Edge & 0.33 & 0.44 & 0.56 & 0.46 \\
(111) Terrace           & 0.46 & 0.52 & 0.50 & 0.48 \\
\hline
\multicolumn{5}{l}{3rd Density Peak} \\
Full Surface            & 0.47 & 0.52 & 0.56 & 0.51 \\
(111)$\times$(111) Edge & 1.74 & 1.59 & 1.23 & 1.38 \\
(111)$\times$(100) Edge & 1.65 & 1.53 & 1.19 & 1.19 \\
\end{tabular}
\end{ruledtabular}
\end{table*}

The profiles in Figure ~\ref{fig:figure2} show that layering of interfacial water extends up to approximately  $\sim$7~\AA{} from the surface. More specifically, the TSD profile is characterised by three distinct peaks (Figures~\ref{fig:figure2}a,b). The first peak, centred at $\sim$2~\AA{}, corresponds to the observation of a chemisorbed first water layer (1\textsuperscript{st} WL), in line with previous AIMD studies~\cite{Khatib2021,Le2020,Sakong2020,Darby2022}. Angular analysis reveals a near-parallel orientation of the water molecule, with its H–O–H bisector tilted slightly away from the surface, consistent with a t\textsubscript{2g}*-like interaction\cite{michaelides2001catalytic}. The integrated 1\textsuperscript{st} WL density yields surface coverage values that increase with potential, in agreement with prior work~\cite{Khatib2021,Raffone2025}. 

A broader and more intense second peak - with a density locally approaching 2.5 g$cm^{-3}$ - 
appears at 3.5 \AA{} (Figures\ref{fig:figure2} b,c), corresponding to the second water layer (2\textsuperscript{nd} WL). In contrast to the more statically chemisorbed first layer (within the simulation timescales explored), the second layer is dynamically structured. The water molecules predominantly adopt H-down orientations, forming a fluctuating hydrogen-bonded network that connects laterally between chemisorbed molecules and extends into the bulk. A weaker third peak is observed at $\sim$6–7 \AA, though—as discussed below—this does not correspond to a genuine third water layer, but instead reflects the spatial extension of the second layer near edge sites.

The CAD profiles (Figure~\ref{fig:figure3}) resolve water structuring around individual step-edge motifs. As in the global TSD analysis, three density maxima are observed, but their intensity and spatial distribution differ across different edge types. At the (111)$\times$(111) edge, the 1\textsuperscript{st} WL peak corresponds to a coverage of $0.93 \pm 0.01$ ML, consistent with full saturation. This high coverage remains stable across all simulated potentials, including values below the PZC. By contrast, the (111)$\times$(100) edge shows a decrease in 1\textsuperscript{st} WL coverage from $0.93 \pm$ to $0.06$ ML at lower potential and to $0.86 \pm 0.04$ ML at higher potential, indicative of partial desorption. This trend reflects a general decrease in oxygen-containing species and is not driven by OH substitution. In both cases, adsorbed water molecules adopt alternating orientations along the edge, with H–O–H bisectors either aligned parallel to the step direction or rotated ~90°, forming directional hydrogen-bonded chains that stabilise the interfacial structure.

A striking feature in the CAD profiles is a cylindrical void of low density ($\sim$100–400~kg m$^{-3}$) above both edges at around 3~\AA{}. Rather than occupying this void, water organises into a broad shell wrapper at 4.5-7.5~\AA{}, which merges with the terrace 2\textsuperscript{nd} WL, while maximising hydrogen bonding around the saturated edge. A minor feature at 3.5–4.5~\AA{} arises from water molecules at the intersection between the cylindrical shell and the terrace water. This becomes more pronounced when the edge is unsaturated, reflecting H-down water transiently occupying the void. As anticipated, the apparent third water layer (3\textsuperscript{rd} WL) in the TSD profile arises primarily from the cylindrical wrapper above edge sites, rather than extended ordering.
The fictitious nature of the third water layer is confirmed by the analysis of the planar-averaged density profiles restricted to the (111) terrace, which show only two distinct water layers, consistent with previous AIMD simulations of flat Pt(111) interfaces~\cite{Sakong2020,Le2020,todorova2024first,Raffone2025}

Water coverage on the (111) top terrace remains consistently lower than at either edge across all simulated potentials, ranging from $0.06\pm0.02$ to $0.09\pm0.02$ ML. Coverage increases linearly with electrode potential, in line with prior AIMD studies \cite{Raffone2025,Khatib2021,Le2020,Sakong2020,Darby2022}. This trend collocates our systems within the intermediate regime of the characteristic sigmoidal coverage–potential relation, bounded by 0.0 and 0.5 ML at low and high potentials, respectively\cite{Raffone2025,Le2021}.
These findings underline the site-specific nature of interfacial water structuring and its sensitivity to local geometry as well as applied bias, with direct implications for surface reactivity and capacitive behaviour at stepped metal electrodes, as will be discussed in the following section.

% CHARGE DISTRIBUTION SECTION
%%%%%%%%%%%%%%%%%%%%%%%%%%%%%%%%%%%%
\subsection{\label{sec:charge distribution}Charge distribution}

%FIGURE4
%%%%%%%%%%%%%%%%%%%%%%%%%%%%%%%%%%%%
\begin{figure}
\includegraphics[width=0.38\textwidth]{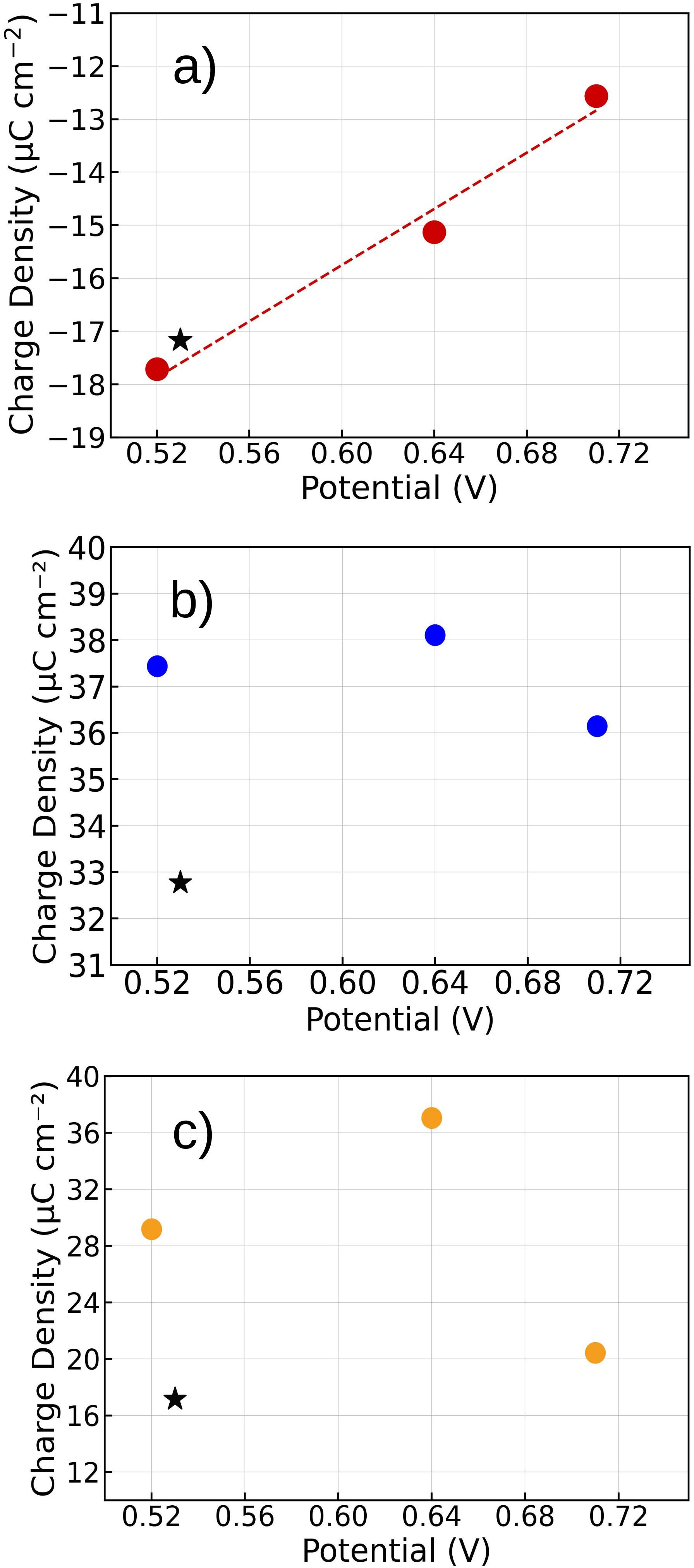}
\caption{\label{fig:figure4}Total charge per area as a function of potential: (a) terrace (111), (b) edge (111)$\times$(111), and (c) edge (111)$\times$(100). The star denotes the potential of zero charge (0.52 V).}
\end{figure}

We show that the surface charge and thus electrode capacitance is primarily governed by the behaviour of water in the 1st chemisorbed WL, which in turn is strongly modulated by surface topology. In line with what observed in other literature\cite{Khatib2021,Le2020,Raffone2025}, each chemisorbed water molecule contributes a consistent net charge of +0.10 \textit{e}, independent of the site (i.e. terrace or edge), however, the local coverage thereof varies significantly; terraces exhibit low, potential-dependent coverage, while edges remain near saturation. This localisation of charge near the edge is reflected in a difference in local potential profiles near the edge, as detailed below.

Bader charge partitioning is used to decompose the total electrode charge $q$ by region: the upper (111) terrace (defined as five Pt rows excluding edge atoms), the (111)$\times$(100) edge, and the (111)$\times$(111) edge (Table~\ref{tab:table2}). The contributions of each region to the differential capacitance of the interface are shown in Figure~\ref{fig:figure4}. The electrode capacitance is obtained from the slope of the charge–potential relation, $dq/dV$, where the total charge $q$ includes both the Bader charges of the metal atoms and the chemisorbed water molecules, and is normalised by the surface area of each region.

Analysis of the charge  localised on the upper terrace, $q_t$, shows a linear increase with potential, remaining net negative across the range studied. The linear capacitive response of the Pt(111) terrace (25.0\,$\mu$C$\cdot$cm$^{-2}$) originates from potential-dependent water adsorption, rather than intrinsic charge redistribution within the metal. On average, each additional water molecule increases the total electrode charge by $+0.33$\,\textit{e}, or 77.9\,$\mu$C$\cdot$cm$^{-2}$. This arises from the combined effect of the charge contribution per chemisorbed water molecule ($+0.10$\,\textit{e}) and the local perturbation of the bonded Pt atom, which becomes more positively charged ($+0.14\pm0.02$\,\textit{e}) compared to bare Pt sites ($-0.09\pm0.00$\,\textit{e}). As the potential increases, more water molecules chemisorb, reducing the population of negatively charged bare Pt atoms and thus increasing the net electrode charge, which becomes less negative.
Interestingly, decomposition of the terrace charge, \( q_t \), reveals linear trends in the charge–potential relation even when only the Bader charge of the metal atoms is considered. The differential capacitance obtained from the metal contribution alone is 19.7\,$\mu$C$\cdot$cm$^{-2}$ for the bare surface, increasing to 25.0\,$\mu$C$\cdot$cm$^{-2}$ when the chemisorbed water layer is included.

Analysis of the (111)$\times$(111) edge reveals that its total charge ($q_{111}$) has no clear dependence on potential (Table \ref{tab:table2} and panel b) in Fig. S2). Decomposition into metal and 1\textsuperscript{st} WL Bader components indicates that both contributions remain nearly constant across the simulated range.   Water coverage over the edge remains high ($0.93 \pm 0.01$~ML), with no adsorption of other species observed. While each chemisorbed water molecule contributes a net charge of $+0.10 \pm 0.01$\,\textit{e}, identical to the terrace case, the bonded Pt atoms carry a lower positive charge ($+0.07 \pm 0.01$\,\textit{e}), approximately half of that of water-bound Pt on the terrace. As the edge is already saturated within the considered potential window, no additional charging occurs via water chemisorption. This indicates that the (111)$\times$(111) edge reaches its maximum water coverage at significantly lower potentials than the terrace and therefore contributes minimally to the differential capacitance around the PZC.

%TABLE2
%%%%%%%%%%%%%%%%%%%%%%%%%%%%%%%%%%%%
\begin{table*}
\caption{\label{tab:table2}Charge per unit area in $\mu\mathrm{C}\,\mathrm{cm}^{-2}$ across the full surface, decomposed into contributions from the (111)$\times$(111) and (111)$\times$(100) step edges, as well as the upper (111) terrace, for systems with different ionic compositions and electrode potentials. Notably, the lower (111) terrace and angled (111)/(100) terraces are not included.}
\begin{ruledtabular}
\begin{tabular}{lc|ccc|ccc|ccc|ccc}
System & Potential &
\multicolumn{3}{c}{Full-surface} &
\multicolumn{3}{c}{(111) $\times$ (111) Edge} &
\multicolumn{3}{c}{(111) $\times$ (100) Edge} &
\multicolumn{3}{c}{(111) Terrace} \\
\hline
 &  & 1st WL & Metal & Total 
 & 1st WL & Metal & Total 
 & 1st WL & Metal & Total 
 & 1st WL & Metal & Total \\
\hline
Water      & 0.53 & 2.36 & -10.94 & -8.58 & 19.18 & 13.59 & 32.77 & 1.07 & 16.09 & 17.16 & 0.98 & -18.14 & -17.17 \\
20H:20F    & 0.52 & 3.10 & -10.87 & -7.77 & 21.97 & 15.47 & 37.44 & 8.29 & 20.87 & 29.17 & 1.22 & -18.93 & -17.71 \\
20H:22F    & 0.64 & 4.37 &  -9.70 & -5.33 & 21.46 & 16.64 & 38.11 & 20.61 & 16.44 & 37.05 & 1.72 & -16.85 & -15.13 \\
18H:22F    & 0.71 & 3.04 &  -8.85 & -5.81 & 21.19 & 14.96 & 36.15 & 1.47 & 18.98 & 20.44 & 2.09 & -14.65 & -12.56 \\
\end{tabular}
\end{ruledtabular}
\end{table*}

Overall, the average charge just outside the (111)$\times$(111) and (111)$\times$(100) step edges is more positive than that on the (111) terrace for all systems studied. %(Table~\ref{tab:potential-drops}).
 For instance, in the reference pure water system, the mean charge per cell at (111)$\times$(111) and (111)$\times$(100) Pt sites is $+0.52\pm0.10$\,e and $+0.61\pm0.08$\,e, respectively, compared to $-3.46\pm0.14$\,e those at the (111) terrace.   
This spatial gradient persists under potential bias: as the potential shifts up from the 20H:20F (-0.01 V) system to the 18H:22F (0.18 V) system, the terrace becomes less negatively charged, namely from $-5.85$ to $-4.76$\,e, while the edge charges remain less affected, confirming that the edges are less prone to response to bias. 

The increase in charge with potential from the (111) terrace and the (111)$\times$(111) edge aligns with experimental measurements on stepped Pt[(111)$\times$(111)] single crystals ($n = 3$–$5$), where the $q$–$V$ response increases linearly from the PZC~\cite{Gomez1999}. The (111)$\times$(100) edge exhibits a stronger charge response to potential than either the terrace or the (111)$\times$(111) edge. This response originates from  adsorbed species, which in this case include OH. By contrast, the metal component of $q_{100}$ does not show systematic dependence on potential. Indeed, unlike the other sites, water decomposition leading to a configuration with adsorbed OH, occurs at this edge, reaching up to 0.12\textbackslash ML. Each OH group carries a net charge of $-0.49\pm0.01$\,\textit{e}, equivalent in magnitude to that of five chemisorbed water molecules but of opposite sign, and thus strongly influences the edge charge. Pt atoms bound to OH carry $+0.13\pm0.03$\,\textit{e}, slightly more than those bound to water ($+0.09\pm0.01$\,\textit{e}), whose value is comparable to that of water-bound Pt on the terrace. 

These results suggest that the (111)$\times$(100) edge is more prone to water dissociation and OH adsorption than other surface motifs, with an OH adsorption window likely shifted to lower potentials relative to the (111)$\times$(111) edge. However, the variation in $q_{100}$ cannot be fully rationalised, as the observed OH coverage likely underestimates equilibrium values due to limited sampling of water dissociation events within the accessible AIMD timescales (30–40\,ps). 

% LATERAL CHARGE DISTRIBUTION SECTION
%%%%%%%%%%%%%%%%%%%%%%%%%%%%%%%%%%%%
\subsection{\label{sec:lateral charge distribution}Lateral Charge Distribution and Step Density Effects}

\begin{figure}
\includegraphics[width=0.46\textwidth]{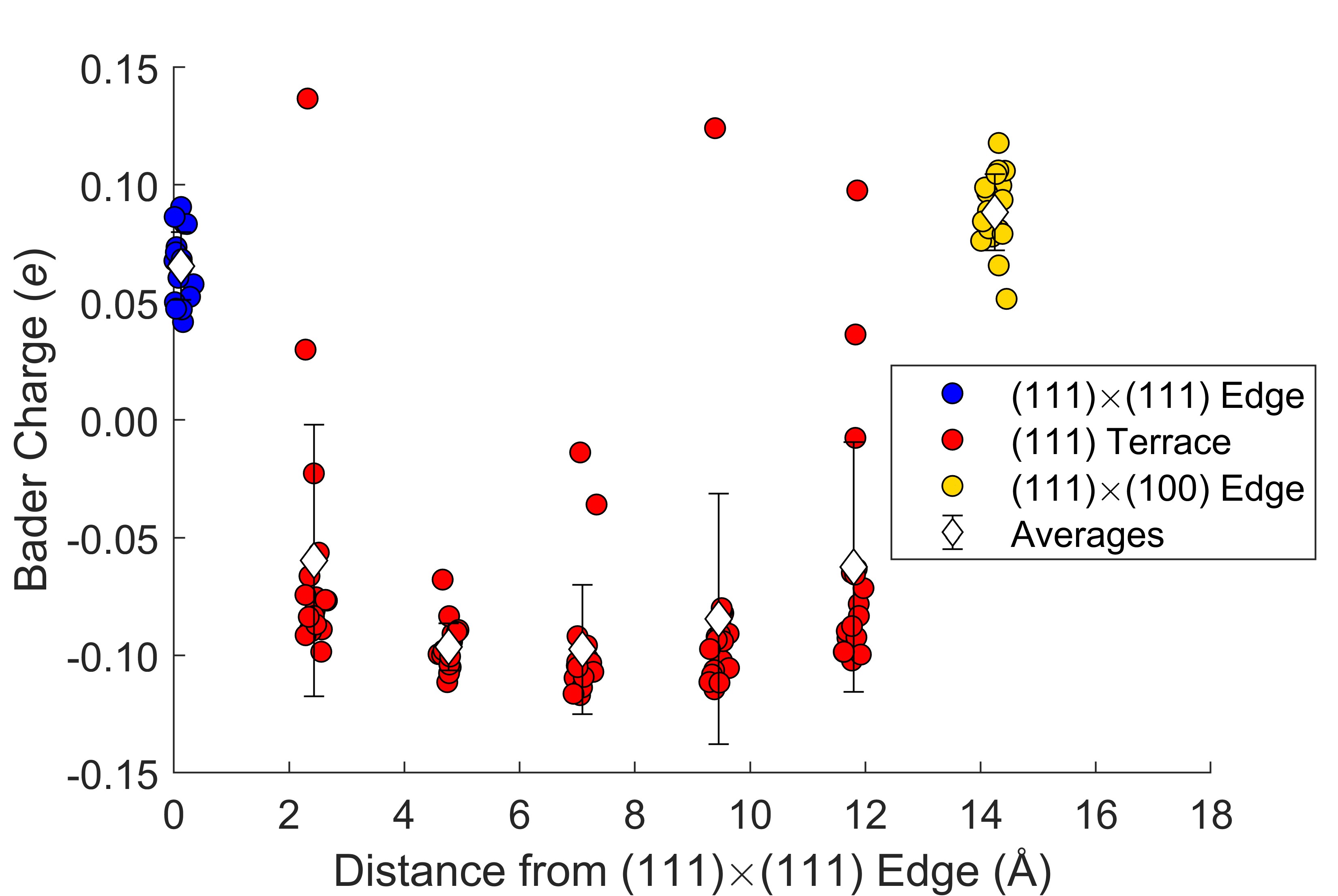}
\caption{\label{fig:figure5}The trajectory average Bader charge per atom as function of the lateral displacement from the average position of the (111)$\times$(111) edge Pt atoms across the terrace for the system at 0.52 V (20H:20F). The charges on (111)$\times$(111) edge atoms (blue circles), (111) terrace atoms (red circles) and (111)$\times$(100) edge atoms (gold circles) are shown, with the distance calculated by considering the displacement from the (111)$\times$(111) edge across the terrace. The average charge for each row of Pt atoms is indicated by diamonds with associated standard deviation.}
\end{figure}

The surface charge distribution is highly heterogeneous across the stepped Pt interface. At all simulated potentials, the (111) terrace is negatively charged, while the undercoordinated edges are positively charged (Table ~\ref{tab:table2}). This polarity gradient is dictated by the differential water coverage and bonding character across surface sites. The lateral charge profile (Figure~\ref{fig:figure5}) shows a sharp inversion from positive to negative charge when moving from the edge to the first terrace row, followed by a gradual decay towards more negative values with increasing terrace depth. 

Thus, it follows that increasing the edge density (i.e., narrowing the terrace) would bias the overall surface charge towards more positive values at fixed potential. This is consistent with experimental trends in the PZC, which decreases linearly with step density for Pt[n(111)$\times$(111)] systems \cite{Gomez1999}. In particular, deviations from the pristine Pt(111) PZC saturate at $-0.18$\,V for terraces $\leq$4 atoms wide (for (111)$\times$(111) steps) and $-0.06$\,V for terraces $\leq$5 atoms wide (for (111)$\times$(100)). Our 7-atom-wide terrace lies within the linear correlation region, supporting that the terrace centre remains representative of an extended Pt(111) surface and that edge interactions are effectively decoupled.

Comparison with the vacuum system confirms that surface charge heterogeneity arises from interfacial polarisation by water. In the absence of solvent, the charge is uniformly negative and lateral charge gradients are negligible.

As previously shown for flat Pt(111), water molecules in the 1\textsuperscript{st} WL donate electron density from O lone pairs to the metal $d$-states, while receiving back-donation into their $t_{2g}$ orbitals. The bonded Pt atom becomes positively charged, while excess negative charge is redistributed laterally across adjacent surface atoms, which become more negative. Analysis of the projected density of states (PDOS) indicates that edge Pt atoms undergo similar interactions.

%FIGURE5
%%%%%%%%%%%%%%%%%%%%%%%%%%%%%%%%%%%%

% LOCAL POTENTIAL DROP AND EDGE REACTIVITY SECTION
%%%%%%%%%%%%%%%%%%%%%%%%%%%%%%%%%%%%
\subsection{\label{sec:local potential drop}Local potential drop and edge reactivity}

\begin{figure*}
\includegraphics[width=0.75\textwidth]{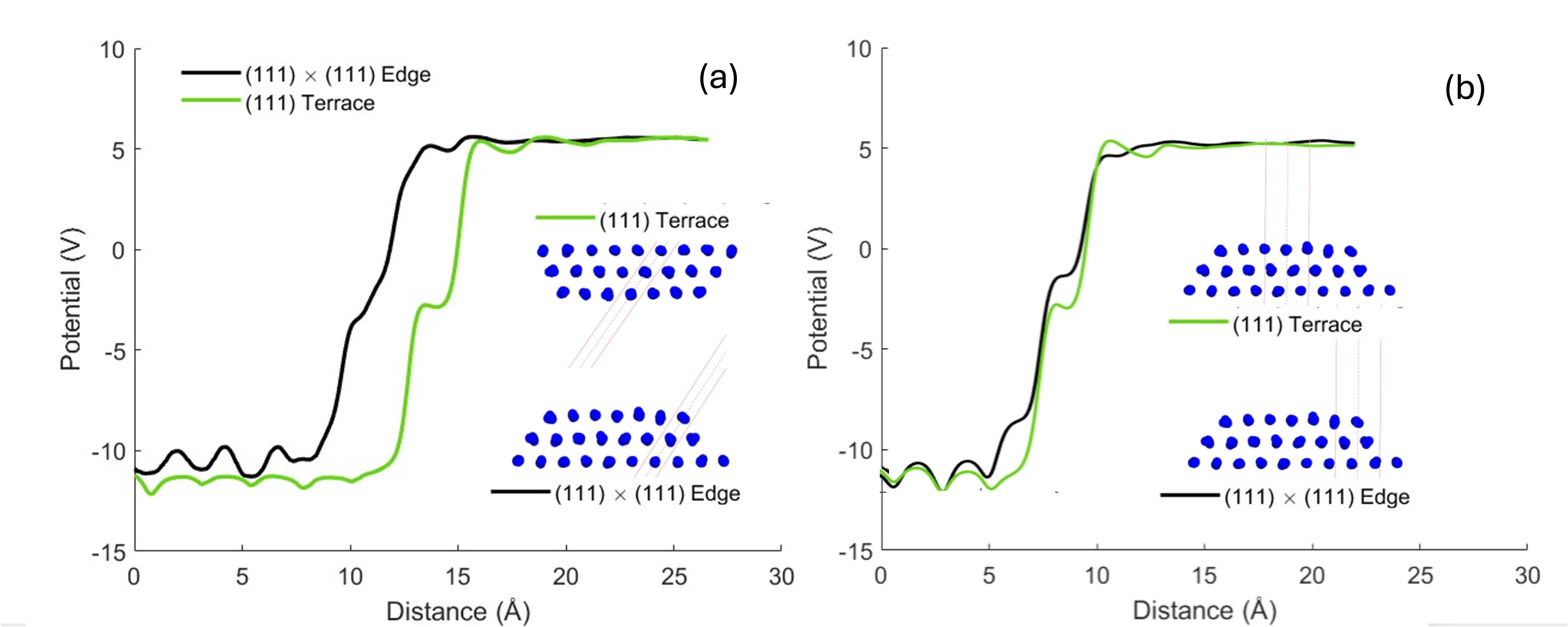}% Here is how to import EPS art
\caption{\label{fig:figure6}Macroscopic electrostatic potential profiles projected from the bulk Pt slab into the electrolyte at two representative surface regions: the (111)$\times$(111) step edge (black) and the (111) terrace (green).  Panel (a) shows the potential profiles obtained along a direction defined by the bisector vector of the (111)$\times$(111) step facet, which intersects the lower terrace near the edge and the upper terrace near its centre. Panel (b) presents the corresponding profiles along the surface-normal direction, aligned with the \( z \)-axis of the simulation cell. In both cases, the electrostatic potential was computed from the planar-averaged Hartree potential and then ''macroscopically averaged'' along the sampling direction indicated in the insets. Averaging was laterally restricted to narrow spatial windows centred on either the step edge or the flat terrace region, as shown in the insets and described in the Supplementary Information. The insets schematically depict the atomic configurations of the sampled metal regions. Red lines indicate the spatial region and direction used for the averaging shown in each graph.}
\end{figure*}

\begin{figure}
\includegraphics[width=0.5\textwidth]{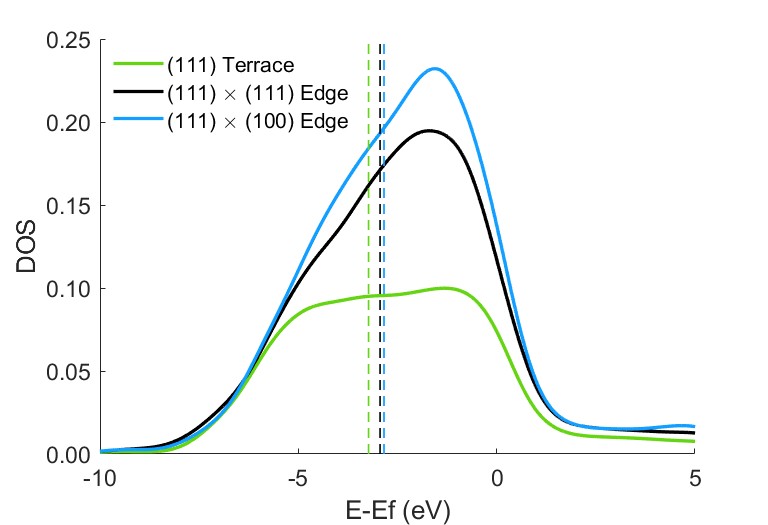}
\caption{\label{fig:figure7}D-band projected density of states (PDOS) of Pt surface atoms in the upper (111) terrace (green), the (111)$\times$(111) edge (black) and the (111)$\times$(100) edge (blue), with the position of the d-band Baricentre indicated by dashed lines, taking values of -3.23, -2.94 abd -2.84 eV, respectively. All energies are given relative to the Fermi level for the system (E\textsubscript{f}).}
\end{figure}

The electrostatic potential profile across the electrified Pt–electrolyte interface reveals pronounced spatial heterogeneity, reflecting the distinct structural and electronic characteristics of step edges and terraces. In both directions of projection—perpendicular to the (111) terrace and along the bisector of the (111)$\times$(111) step facet—the edge region exhibits a consistently higher electrostatic potential compared to the terrace (Figure~\ref{fig:figure6}). This implies that an approaching electron would experience a more repulsive (i.e., less favourable) potential near the edge, consistent with its role as a locus of positive surface charge. The profiles shown are laterally averaged over spatial windows that isolate the step and terrace regions individually; full methodological details are provided in the Supplementary Information. The surface data to obtain these profiles are laterally restricted such that they include only contributions from the terrace and edge atoms, respectively (See SI). Important insight into edge local potential reactivity is provided by analysis of the PDOS. As shown in Table~\ref{tab:table3} and Figure~\ref{fig:figure7}, the baricentre, or first moment, of the d-band for edge atoms ((111)$\times$(111): –2.94 eV; (111)$\times$(100): –2.84eV) lies consistently higher in energy (closer to the Fermi level) than for (111) terrace atoms (–3.23 eV), indicating a higher density of metal states available for bonding. According to d-band theory,
%\cite{Norskov2004} 
\cite{doi:10.1021/jp047349j} this upward shift can be related to a stronger overlap with adsorbate anti-bonding orbitals, and thus enhanced chemical reactivity at step sites.

%FIGURE6
%%%%%%%%%%%%%%%%%%%%%%%%%%%%%%%%%%%%

Additionally, the edge-localised PDOS profiles are notably narrower and more sharply peaked than their terrace counterparts, suggesting a higher effective electronic mass and stronger spatial confinement of the d-states. This localisation facilitates stronger coupling to nearby interfacial dipoles (such as chemisorbed water and OH) and enhances the local field response, in line with the elevated electrostatic potentials observed at edges.\\
% (Table~\ref{tab:potential-drops}). \\

%TABLE3
%%%%%%%%%%%%%%%%%%%%%%%%%%%%%%%%%%%%
\begin{table}
\caption{\label{tab:table3}First moment of the d-band projected density of states (PDOS) for selected surface regions. Values are given in eV relative to the Fermi level.}
\begin{ruledtabular}
\begin{tabular}{lc}
Region & d-band CoM (eV) \\
\hline
Upper (111) terrace     & –3.23 \\
(111)$\times$(111) edge & –2.94 \\
(111)$\times$(100) edge & –2.84 \\
All surface             & –2.94 \\
\end{tabular}
\end{ruledtabular}
\end{table}

These findings align with experimental measurements on stepped Pt[n(111)$\times$(111)] single crystals, where increasing step density lowers the PZC and modifies the interfacial charge–voltage response.\cite{Gomez1999} A lower PZC implies that a more negative electrode potential is required to neutralise the surface, consistent with an increased equilibrium accumulation of positive charge at step edges compared to flat Pt(111).

Overall, these electronic signatures demonstrate that step edges act as focal points of charge accumulation, enhanced field, and potential catalytic activity.\\

%FIGURE7
%%%%%%%%%%%%%%%%%%%%%%%%%%%%%%%%%%%%

% CONCLUSION
%%%%%%%%%%%%%%%%%%%%%%%%%%%%%%%%%%%%
\section{\label{sec:conclusion}Conclusion}
This work establishes a molecular-level framework for understanding the structure and capacitive behaviour of the electric double layer (EDL) at realistic, stepped Pt–water interfaces. Using \textit{ab initio} molecular dynamics at controlled potentials on a model incorporating both (111)$\times$(111) and (111)$\times$(100) edge motifs, we reveal how nanoscale surface topography governs interfacial charge storage, electrostatic potential profiles, and local chemical reactivity.

We find that differential capacitance near the potential of zero charge (PZC) arises primarily from potential-dependent adsorption of charged water molecules on flat (111) terraces. Water coverage on terraces increases linearly with bias, while step edges contribute negligibly to the capacitance due to saturation with chemisorbed water below the PZC. This asymmetry reflects locally higher electrostatic potentials at step sites, which stabilise positive surface charge and inhibit further adsorption.

EDL structure across the interface exhibits densification into a bilayer architecture. The first layer consists of flat-lying chemisorbed water molecules that each contribute approximately 0.1\,$|e|$ to the surface charge. On the terraces, this layer grows with increasing potential; at the edges, it saturates early and is further stabilised through zigzag hydrogen-bonding networks connected to the second water layer. The second layer displays predominantly H-down orientation across the surface and wraps cylindrically around step edges, enclosing low-density voids above the saturated chemisorbed shell.

Local electronic structure analysis confirms that step atoms are more electropositive than terrace atoms. Their d-band projected density of states (PDOS) is both higher in energy and more sharply defined, consistent with enhanced reactivity and stronger interaction with interfacial dipoles. These observations align with d-band theory predictions and support a mechanism in which step sites act as stabilised centres of charge accumulation.

The suppressed capacitive response of step edges, driven by higher in energy local potentials and early saturation, leads to a shift in the overall charging behaviour of the electrode. As a result, stepped Pt surfaces require more negative potentials to reach net neutrality. This mechanistic insight rationalises the experimentally observed lowering of the PZC with increasing step density on Pt[n(111)$\times$(111)] single crystals.\cite{Gomez1999}

Together, these findings provide an atomistically resolved and mechanistically grounded explanation of how surface morphology shapes electric double layer formation, charge localisation, and interfacial reactivity. This work advances the theoretical understanding of platinum nanostructuring strategies and informs the rational design of high-performance electrocatalytic interfaces.

% ACKNOWLEDGEMENT
%%%%%%%%%%%%%%%%%%%%%%%%%%%%%%%%%%%%
\begin{acknowledgments}
We are thankful for the useful discussions with Dr. Alexandre Lozovoi.
This work was supported by the Engineering and Physical Sciences
Research Council (grant EP/P033555/1). We also gratefully acknowledge the use of the High-Performance Computers at Imperial College London, provided by Imperial College Research Computing Service DOI: 10.14469/hpc/2232, and the computing resources provided by STFC Scientific Computing Department’s SCARF cluster. The authors gratefully acknowledge the ARCHER2 UK National Supercomputing Service (https://www.archer2.ac.uk), via our membership of the UK’s HEC Materials Chemistry Consortium, which is funded by EPSRC (EP/R029431 and EP/X035859). The authors gratefully acknowledge the Gauss
Centre for Supercomputing e.V. (www.gauss-centre.eu) for funding this
projectbyproviding computing timeon theGCSSupercomputerSuperMUCNG
at Leibniz Supercomputing Centre (www.lrz.de). MSi  acknowledges funding by the German Research Foundation Germany’s Excellence Strategy-EXC 2033-390677874-RESOLV.
Finally, we would like to acknowledge the Thomas Young Centre under grant number TYC-101.
Author Contributions: CSC conceived and supervised the research. MD performed the calculations and provided a first draft for the paper, finalised by CSC. All authors (MSh, MD, MSi and CSC) conducted and discussed the analyses. All authors proofread the manuscript.
\end{acknowledgments}

\section{Competing Interests}
The authors declare that they have no competing interests.

\section*{Data Availability Statement}

All data needed to evaluate the conclusions in the paper are present in the paper and/or the Supplementary Information. Additional data related to this paper may be requested from the authors.

\bibliography{reference}% Produces the bibliography via BibTeX.

\end{document}